\documentclass[]{spie}  %>>> use for US letter paper

\usepackage{amsmath,amsfonts,amssymb}
\usepackage{graphicx}
\usepackage{subcaption}
\usepackage{setspace}
\usepackage{tocloft}
\usepackage{booktabs}
\usepackage{placeins}
\usepackage{float}
\usepackage{listings} 
\usepackage{siunitx}

\usepackage[ruled,noline,linesnumbered]{algorithm2e}

\title{Deep Learning Framework for Digital Breast Tomosynthesis Reconstruction}

\author[a]{Nikita Moriakov}
\author[a]{Koen Michielsen}
\author[c, d]{Jonas Adler}
\author[a]{Ritse Mann}
\author[a, e]{Ioannis Sechopoulos}
\author[a, b]{Jonas Teuwen}

\affil[a]{Radboud University Medical Center, Diagnostic Image Analysis Group, Department of Radiology and Nuclear Medicine, Nijmegen, the Netherlands}
\affil[b]{Optics Research Group, Imaging Physics Department, Delft University of Technology, the Netherlands}
\affil[c]{Department of Mathematics, KTH Royal Institute of Technology, Stockholm, Sweden}
\affil[d]{Research and Physics, Elekta, Stockholm, Sweden}
\affil[e]{Dutch Expert Centre for Screening, Nijmegen, the Netherlands}

\authorinfo{Send correspondence to Nikita Moriakov: nikita.moriakov@radboudumc.nl.}

\cftpagenumbersoff{figure}
\cftpagenumbersoff{table} 
\begin{document} 
\maketitle

\begin{abstract}
Digital breast tomosynthesis is rapidly replacing digital mammography as the basic x-ray technique for evaluation of the breasts. However, the sparse sampling and limited angular range gives rise to different artifacts, which manufacturers try to solve in several ways. In this study we propose an extension of the Learned Primal-Dual algorithm for digital breast tomosynthesis. The Learned Primal-Dual algorithm is a deep neural network consisting of several `reconstruction blocks', which take in raw sinogram data as the initial input, perform a forward and a backward pass by taking projections and back-projections, and use a convolutional neural network to produce an intermediate reconstruction result which is then improved further by the successive reconstruction block. We extend the architecture by providing breast thickness measurements as a mask to the neural network and allow it to learn how to use this thickness mask. We have trained the algorithm on digital phantoms and the corresponding noise-free/noisy projections, and then tested the algorithm on digital phantoms for varying level of noise. Reconstruction performance of the algorithms was compared visually, using MSE loss and Structural Similarity Index. Results indicate that the proposed algorithm outperforms the baseline iterative reconstruction algorithm in terms of reconstruction quality for both breast edges and internal structures and is robust to noise. 
\end{abstract}

% Include a list of up to six keywords after the abstract

% {\noindent \footnotesize\textbf{*}Equal contribution}\\
% Include email contact information for corresponding author
%{\noindent \footnotesize\textbf{\dag}Corresponding author,  \linkable{jonas.teuwen@radboudumc.nl} 
%}

\begin{spacing}{1}   % use double spacing for rest of manuscript

\section{Introduction}
\label{sect:intro}  % \label{} allows reference to this section
Digital breast tomosynthesis (DBT) is rapidly replacing digital mammography (DM) as the basic x-ray technique for evaluation of the breasts. DBT overcomes some of the inherent limitations of DM by adding limited depth information to mammographic images. This prevents inherent information loss caused by tissue superposition, and may even increase specificity by resolving tissue projections that mimick breast lesions. A DBT acquisition consists of several low-dose planar x-ray projections at equally spaced intervals over a limited angle. These projections are then reconstructed to a three-dimensional volume. However, this sparse sampling and limited angular range gives rise to different artifacts, which manufacturers try to solve in several ways. In previous work \cite{rodriguez2017} it was shown that the chosen reconstruction algorithm can greatly influence the reconstruction quality.

In this paper, we are interested in DBT reconstruction with a data-driven approach using deep learning. As we will show, not only does this allow us to easily include complicated (shape) priors into the reconstruction, but it additionally opens opportunities for end-to-end learning and predicting lesion locations in CAD-systems, or to compute the accumulated x-ray dose. We propose a data-driven reconstruction algorithm, Deep Breast Tomographic Reconstruction (DBToR) using a deep neural network, which extends the previously proposed Learned Primal-Dual algorithm. The neural network consists of several `reconstruction blocks', which take in raw sinogram (i.e., projection) data as the initial input, perform a forward and a backward pass by taking projections and back-projections, and use a convolutional neural network to produce an intermediate reconstruction result which is then improved further by each successive reconstruction block. This neural network is trained by stochastic gradient descent, which minimizes the $L^2$ loss between the reconstruction computed reconstruction algorithms to determine the reconstruction volume. In contrast to CT imaging, measurements from only a narrow range of sparsely sampled angles are available in breast tomosynthesis. However we also have access to information on the compressed breast thickness, and it is used in the classical reconstruction algorithms to determine the reconstruction volume. We provide these thickness measurements as additional prior information to the Learned Primal-Dual algorithm by giving it a mask and allow it to learn how to use this mask efficiently. 

We have tested the algorithm on virtual breast phantoms. The results indicate that the proposed algorithm outperforms the baseline iterative reconstruction algorithm in terms of reconstruction quality for both breast edges and breast internal structures. Furthermore, the algorithm generalizes well even when trained on a small dataset and is robust to noise.

\section{Methods}
\subsection{Material}
To train and evaluate the algorithm, we created a total of 1124 simulated breast phantoms. To limit computational complexity, the phantoms consisted of 2D coronal slices extracted from virtual 3D breast phantoms~\cite{Lau2012-uc}. 
These phantoms were indexed with labels for four different materials: skin, adipose tissue, glandular tissue, and Cooper's ligaments. The elemental compositions of these materials were obtained from the work of Hammerstein et. al.\cite{Hammerstein1979-uc}, except for the composition of Cooper's ligaments, which was assumed to be identical to that of glandular tissue. 
Linear attenuation coefficients at \SI{20}{\kilo\electronvolt} were calculated for each material using the software from Boone and Chavez~\cite{Boone1996-tu}. 
The phantoms include compressed breast thicknesses from \SIrange{3.0}{5.6}{\centi\meter} and widths from \SIrange{5.8}{18.0}{\centi\meter} with an isotropic voxel size of \SI{0.2}{\milli\metre}$\times$\SI{0.2}{\milli\metre}.

\begin{figure}[h!]
\centering
\includegraphics[width=0.3\textwidth]{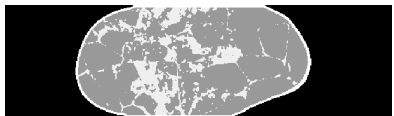}
\caption{Sample breast image}
\label{fig:gt}
\end{figure}
\vspace{-0.3cm}
%### Projection data ###
Limited angle fan-beam projections were simulated for all phantoms using a geometry with the center of rotation placed at the bottom center of the phantom. The x-ray source was placed \SI{650}{\milli\metre} above the center of rotation, and the source-detector distance was \SI{700}{\milli\metre}. A total of 25 equally spaced projections between \SIlist{-24;24}{\degree} were generated, with the detector rotating with the x-ray source. The detector was a perfect photon counting system consisting of 1280 elements of \SI{0.2}{\milli\metre} width. The forward model $\hat{y}_i (\vec{x}) = b_i e^{- \sum_j l_{ij} x_j}$
was used for the simulations, with $\hat{y}$ the simulated projection data, $b_i$~the number of x-ray photons emitted towards detector pixel~$i$, $l_{ij}$~the intersection between voxel~$j$ and the line between the source and detector pixel~$i$, and $x_j$ the linear attenuation in voxel~$j$.
The noiseless simulated projection data were used to generate a series of data sets at 17 noise levels. This was simulated by setting photon count $b_i = 1000 \cdot \sqrt{2}^N$ with $N = 4, 8, 12$. The cases with $N=8$ have a noise level of similar magnitude to clinical DBT projection data. For each noise level, 10 Poisson noise-realizations were generated, resulting in a total of 11240 projection sets at each dose level.

Reference reconstructions were generated for both noiseless and noisy data using 100 iterations of MLTR without any regularization~\cite{Nuyts1998-ea}.

\subsection{Algorithm}
The DBToR algorithm, which we propose for the problem is a modification of the Learned Primal-Dual Algorithm\cite{lpdr} (LPD), which we extend by taking breast thickness measurements into account in order to improve reconstruction quality. These breast thickness measurements are computed as the distance between the detector cover plate and the compression paddle, and are available during testing. The measurements of breast thickness for breast $x$ can be turned into a 2D mask $\mathrm{height\_mask}_x$, which is a mask with constant height and full-width, which restricts the region in which the breast is located in one axis. Compared to the base LPD algorithm, we have seen that the addition of mask information leads to more stable training and higher reconstruction quality. Complete algorithm training procedure is provided as Algorithm \ref{algo:recon}, where $\mathcal P$ is the projection operator and $\mathcal P^*$ is the backprojection. At test time, we compute the reconstruction from the given height mask $m$ and the projections $p$ as $\mathrm{compute\_reconstruction}(m, p)$.

\begin{figure}
\begin{algorithm}[H]
\small
\caption{DBToR algorithm and training}\label{algo:recon}
\DontPrintSemicolon
 \SetKwFunction{comprec}{compute\_reconstruction}
 \SetKwProg{Fn}{Function}{:}{}
 \Fn{\comprec{$m$, $p$}}{
        $f_0 \leftarrow 0 \in X^{N_{prim}}$; \emph{\# Initialize primal vector}\
        
        $h_0 \leftarrow 0 \in U^{N_{dual}}$; \emph{\# Initialize dual vector}\
        
        \For{$i \leftarrow 0$ \KwTo $I$}{
            $h_i \leftarrow \Gamma_{\theta_i^d} (h_{i-1}, \ \mathcal P(f_{i-1}^{(2)}), \ p,  \ \mathcal P(m)))$;\
            
            $f_i \leftarrow \Lambda_{\theta_i^p}(f_{i-1}, \ \mathcal P^*(h_i^{(1)}), \ m)$;\
        }
        \KwRet $f_{I}^{(1)}$\;
 }
  
\For{$j \leftarrow 0$ \KwTo $N_{\text{iter}} - 1$}{
\uIf{$j \% \mathrm{freq} == 0$}{

$x \sim \mathcal D_{train}$;  \emph{\# Sample image from the training dataset}\ 

$y \leftarrow \mathrm{input\_sinograms}_x$; \emph{\# Retrieve corresponding projection data}\

$m \leftarrow \mathrm{height\_mask}_x$; \emph{\# Retrieve corresponding mask data}\
}

$z \leftarrow \mathrm{compute\_reconstruction}(m, y)$;\

$loss \leftarrow \| z - x \|_2^2$;\

change parameters $\theta_i^p, \theta_i^d, i=1, \dots, I$ to reduce $loss$;
}
\end{algorithm}
\vspace{-0.5cm}
\end{figure}

Input sinogram data is log-transformed, after which we scale it so that the resulting mean and standard deviation across the training dataset equal $1$. Forward projection is a linear operator. In all experiments $I=10, N_{prim} = N_{dual} = 5$ and freq=1. $N_{iter} = 4 \cdot 10^5$ for training on noisy data and $N_{iter} = 10^5$ for training on noise-free data. $\Gamma_{\theta_i^d}, \ \Lambda_{\theta_i^p}$ for $i=1, \dots, I$ are neural networks with weights $\theta_i^p, \theta_i^d$ respectively, which we call the \emph{dual reconstruction block} and the \emph{primal reconstruction block}. Each primal/dual reconstruction block is a ResNet-type block consisting of 3 convolutional layers, which is similar to the reconstruction blocks in LPD algorithm~\cite{lpdr}. We initialize $h_0, f_0$ by zeros. Parameters of the neural network are optimized by performing an iteration of ADAM optimizer (line 16) using cosine annealing as a learning rate schedule starting at a learning rate starting of $10^{-3}$.
\begin{figure}[h!]
        \centering
        \parbox{.8\textwidth}{
            \begin{subfigure}{.3\linewidth}
                \includegraphics[width=\textwidth]{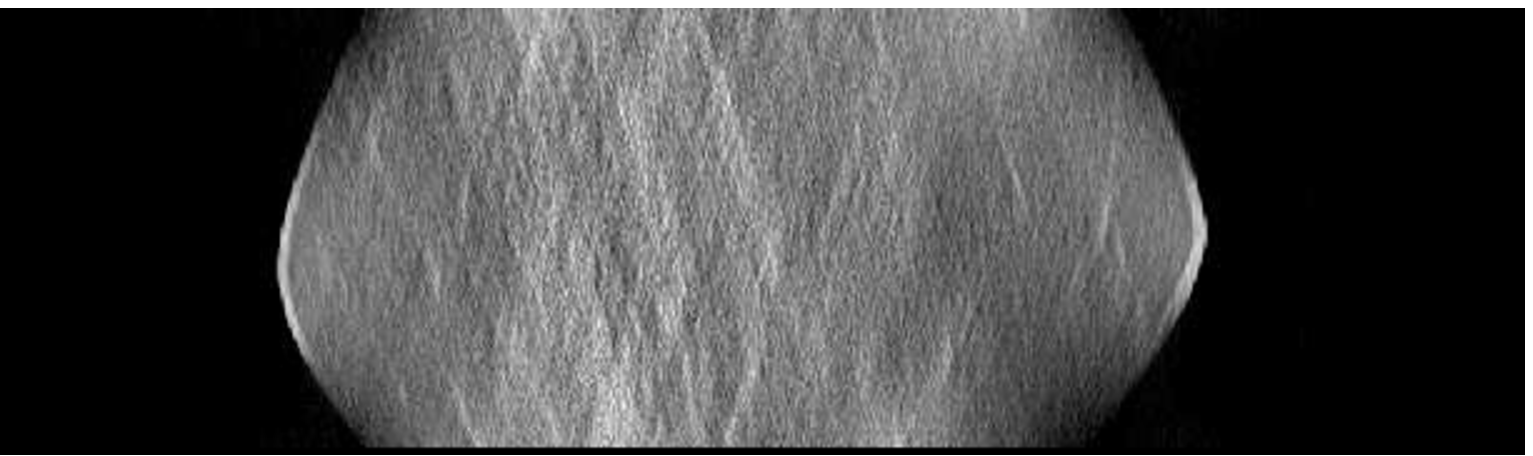}
                \caption{Baseline, $N=4$}
        \end{subfigure}
        \begin{subfigure}{.3\linewidth}
                \includegraphics[width=\textwidth]{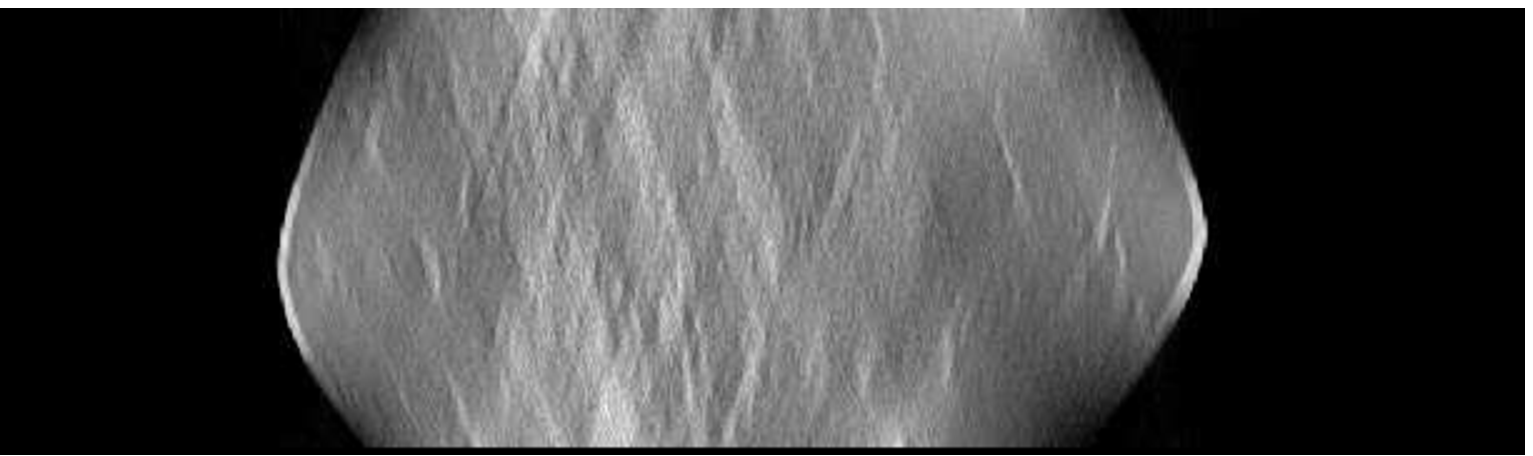}
                \caption{Baseline, $N=8$}
        \end{subfigure}
        \begin{subfigure}{.3\linewidth}
                \includegraphics[width=\textwidth]{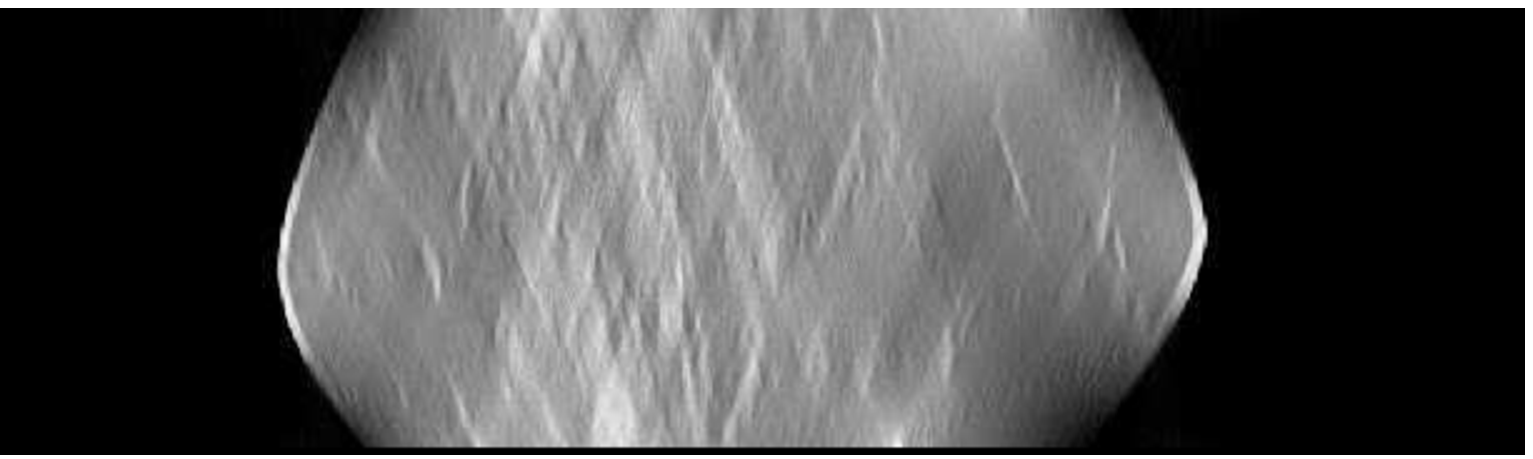}
                \caption{Baseline, $N=12$}
        \end{subfigure}\\
        \begin{subfigure}{.3\linewidth}
                \includegraphics[width=\textwidth]{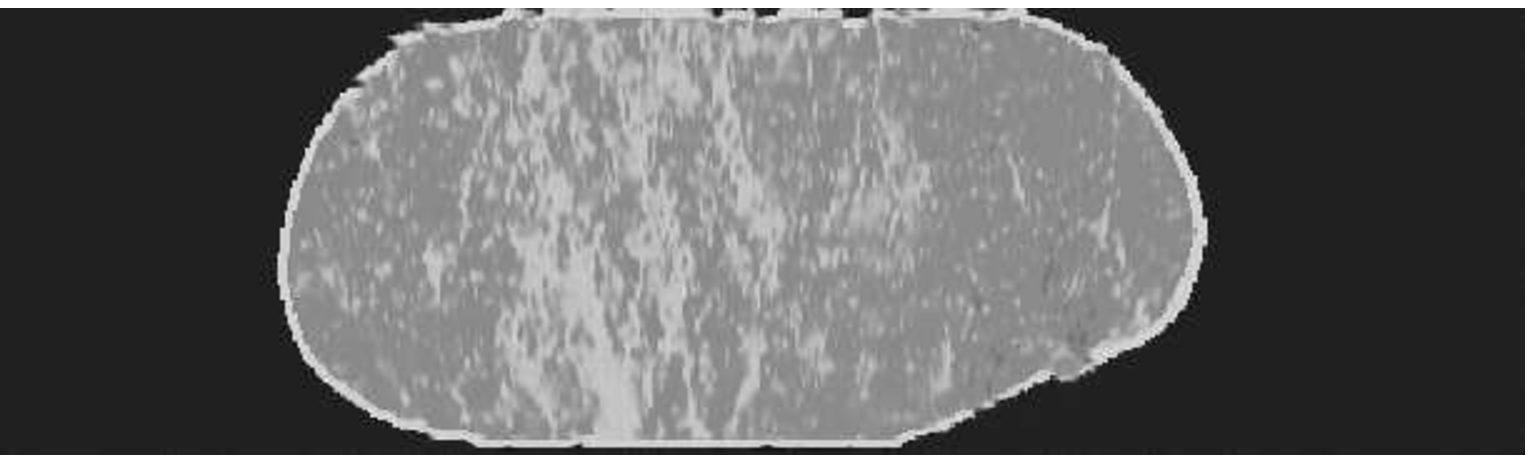}
                \caption{DBToR, $N=4$}
        \end{subfigure}
        \begin{subfigure}{.3\linewidth}
                \includegraphics[width=\textwidth]{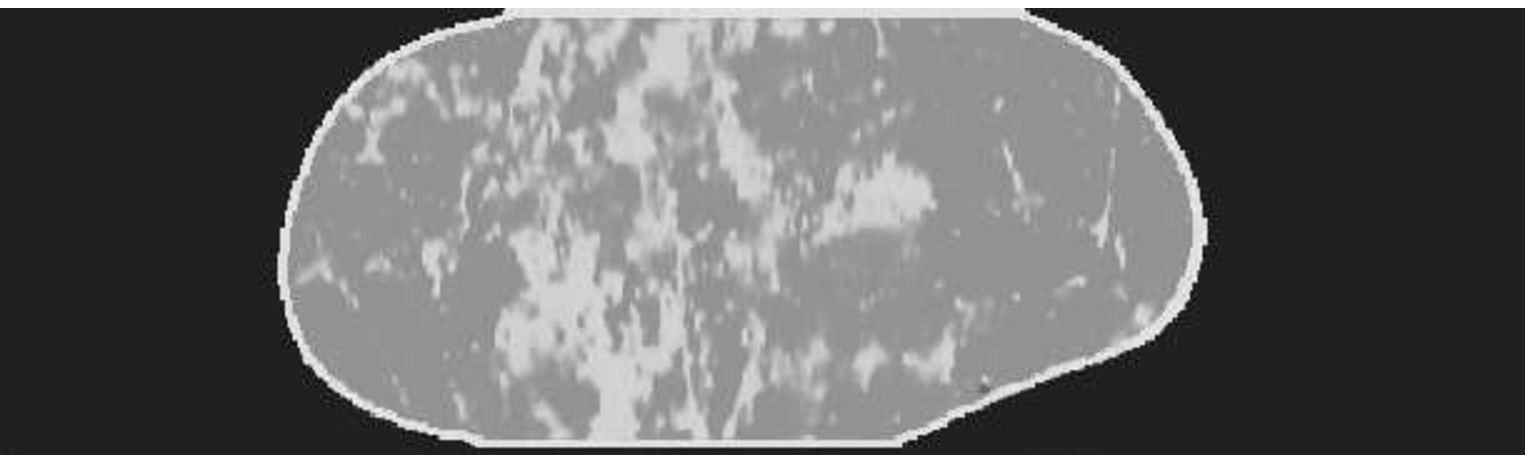}
                \caption{DBToR, $N=8$}
        \end{subfigure}
        \begin{subfigure}{.3\linewidth}
                \includegraphics[width=\textwidth]{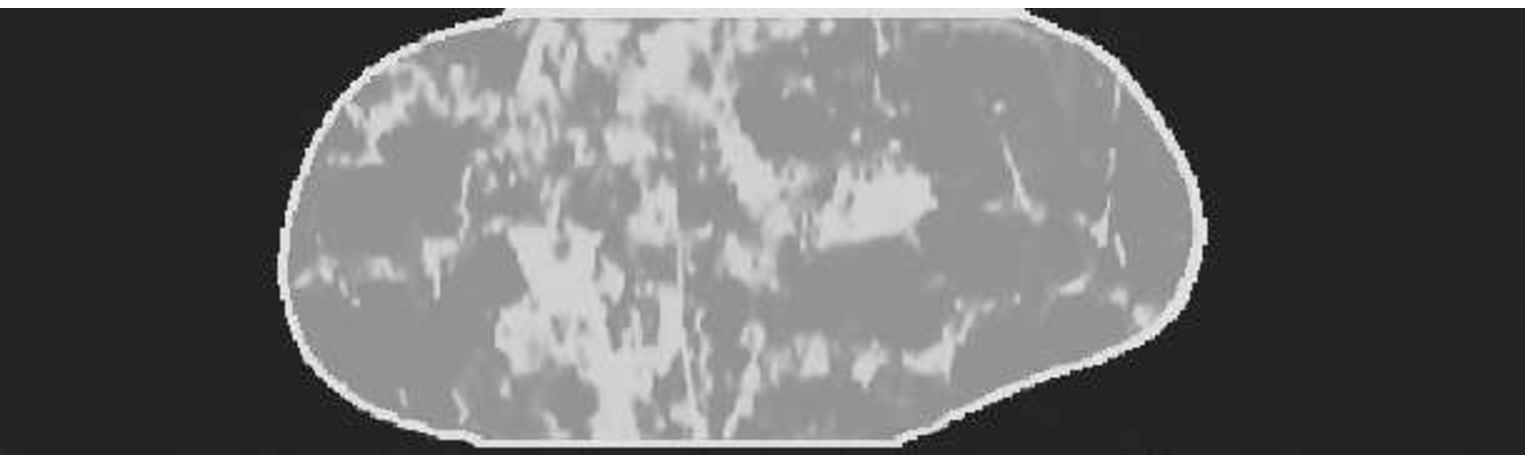}
                \caption{DBToR, $N=12$}
        \end{subfigure}
        }
        \caption{Sample reconstructions for different noise levels}\label{reduced}
\label{samplerec}
\end{figure}
\vspace{-0.5cm}

\section{Results}
In this section we provide a summary of the results and compare the proposed DBToR algorithm to the baseline iterative reconstruction algorithm and the Learned Primal-Dual algorithm. We trained two versions of the DBToR algorithm: one on noise-free projections and one on noisy projections at noise level $N=8$.
For DBToR trained on noise-free data we report the corresponding $L^2$ loss, Structural Similarity Index (SSIM) and Peak Signal-to-Noise Ratio (PSNR) on noise-free test data in Table \ref{tab1}, 
while for DBToR trained on noisy projections we report these metrics for noise levels $N=4, 8, 12$ in Table \ref{tab2}. These results were obtained by making 3 random cross-validation splits with approximately 50\% for training and 50\% for testing at the `patient' level, thus we ensured that all breast slices for any specific patient belong to either the train or to the test set for each split. For noise-free projections, we trained the basic LPD algorithm in addition to DBToR in order to compare the performance (Table \ref{tab1}). Since LPD performed poorly on noise-free projections, we excluded it from further training on noisy projections. 

The proposed DBToR algorithm outperforms the baseline iterative reconstruction algorithm at all noise levels and for all metrics being considered, while yielding visually more accurate reconstructions as well (see ground truth in Figure \ref{fig:gt} and reconstructions in Figure \ref{samplerec}). The LPD algorithm is significantly outperformed in the noise-free case. It is also interesting to note from Table \ref{tab2} that the performance of DBToR at noise level $N=4$ is comparable to the baseline iterative reconstruction algorithm at noise level $N=8$, which corresponds to a 4 times higher photon count. Further performance gains are to be expected when training on a larger dataset.

\begin{table}
\small
\centering
\caption{Result summary for testing on noise-free projections, mean $\pm$ standard deviation across 3 cross-validation dataset splits are given for each algorithm and each metric.}\label{tab1}
\begin{tabular}{llll}
\toprule
Model & $L^2$-loss & SSIM & PSNR \\
\midrule
Baseline (noise-free) &  $0.00773 \pm 4.3 \cdot 10^{-5}$ & $0.83 \pm 4.2 \cdot 10^{-3}$ & $20.19 \pm 2.9 \cdot 10^{-2}$ \\
LPD algorithm (noise-free) & $0.038 \pm 3.8 \cdot 10^{-2}$ & $0.38 \pm 1.8 \cdot 10^{-1}$ & $16.1 \pm 5.7$ \\
DBToR algorithm (noise-free) & $0.0024 \pm 3.0 \cdot 10^{-4}$ & $0.89 \pm 1.0 \cdot 10^{-2}$ & $25.8 \pm 5.5 \cdot 10^{-1}$\\
\bottomrule
\end{tabular}
\end{table}

\begin{table}
\small
\centering
\caption{Result summary for testing on noisy projections for different noise levels $N$, mean $\pm$ standard deviation across 3 cross-validation dataset splits are given for each algorith, metric and noise level.}\label{tab2}
\begin{tabular}{llll}
\toprule
Model & $L^2$-loss & SSIM & PSNR \\
\midrule
Baseline ($N=4$) & $0.0096 \pm 1.0 \cdot 10^{-4}$ & $0.6788 \pm 9 \cdot 10^{-3}$ & $19.17 \pm 5.8 \cdot 10^{-2}$ \\
DBToR ($N=4$) & $0.0075 \pm 2.3 \cdot 10^{-3}$ & $0.7528 \pm 6.4 \cdot 10^{-2}$ & $20.78 \pm 1.18$ \\
\midrule
Baseline ($N=8$) & $0.0082 \pm 5.2 \cdot 10^{-5}$ & $0.7463 \pm 7.2 \cdot 10^{-3}$ & $19.87 \pm 3.6 \cdot 10^{-2}$ \\
DBToR ($N=8$) & $0.0021 \pm 9.7 \cdot 10^{-5}$ & $0.8612 \pm 6.0 \cdot 10^{-2}$ & $26.54 \pm 2.5 \cdot 10^{-1}$ \\
\midrule
Baseline ($N=12$) & $0.0078 \pm 4.47 \cdot 10^{-5}$ & $0.7977 \pm 5.5 \cdot 10^{-3}$ & $20.06 \pm 3.0 \cdot 10^{-2}$ \\
DBToR ($N=12$) & $0.0019 \pm 1.3 \cdot 10^{-4}$ & $0.8689 \pm 6.2 \cdot 10^{-2}$ & $26.62 \pm 2.8 \cdot 10^{-1}$ \\
\bottomrule
\end{tabular}
\end{table}

\section{Discussion and conclusions}
We have presented DBToR, a modification of the Learned Primal-Dual reconstruction algorithm, which is specifically suited for digital breast tomosynthesis. We showed that adding priors such as the breast thickness improves learning stability, generalization and reconstruction quality. Furthermore, we have shown that the DBToR algorithm outperforms the baseline iterative reconstruction algorithm and is robust to noise. 

This paper has not been submitted for consideration elsewhere.

\bibliographystyle{spiejour}   % makes bibtex use spiejour.bst

\end{spacing}
\end{document}